\documentclass[pra,showpacs,twocolumn,superscriptaddress,nofootinbib]{revtex4-1}
\usepackage{graphicx}
\usepackage{dcolumn}
\usepackage{amsmath}
\usepackage{epsfig}
\usepackage{bm}
\usepackage{braket}
\usepackage{bbold}
\usepackage{bbm}
\usepackage{amssymb}
\usepackage[]{hyperref} 
\hypersetup{colorlinks=true,linkcolor=blue,citecolor=blue,urlcolor=black}

\begin{document}

\title{Characterizing the local vectorial electric field near an atom chip using Rydberg state spectroscopy}

\author{N. Cisternas}
\email{n.v.cisternassanmartin@uva.nl}
\affiliation{Van der Waals-Zeeman Institute, Institute of Physics, University of Amsterdam, The Netherlands}

\author{Julius de Hond}
\affiliation{Van der Waals-Zeeman Institute, Institute of Physics, University of Amsterdam, The Netherlands}

\author{G. Lochead}
\email{Present address: Physikalisches Institut, Universitat Heidelberg, Germany}
\affiliation{Van der Waals-Zeeman Institute, Institute of Physics, University of Amsterdam, The Netherlands}

\author{R.J.C. Spreeuw}
\affiliation{Van der Waals-Zeeman Institute, Institute of Physics, University of Amsterdam, The Netherlands}

\author{H.B. van Linden van den Heuvell}
\affiliation{Van der Waals-Zeeman Institute, Institute of Physics, University of Amsterdam, The Netherlands}

\author{N.J. van Druten}
\affiliation{Van der Waals-Zeeman Institute, Institute of Physics, University of Amsterdam, The Netherlands}

\date{\today}

\begin{abstract}

We use the sensitive response to electric fields of Rydberg atoms to characterize all three vector components of the local electric field close to an atom-chip surface. We measured Stark-Zeeman maps of $S$ and $D$ Rydberg states using an elongated cloud of ultracold Rubidium atoms ($T\sim2.5$ $\mu$K) trapped magnetically $100$ $\mu$m from the chip surface. The spectroscopy of $S$ states yields a calibration for the generated local electric field at the position of the atoms. The values for different components of the field are extracted from the more complex response of $D$ states to the combined electric and magnetic fields. From the analysis we find residual fields in the two uncompensated directions of $0.0\pm0.2$ V/cm and $1.98\pm0.09$ V/cm respectively. This method also allows us to extract a value for the relevant field gradient along the long axis of the cloud. The manipulation of electric fields and the magnetic trapping are both done using on-chip wires, making this setup a promising candidate to observe Rydberg-mediated interactions on a chip.
\end{abstract}

\date{\today}


\maketitle


\section{Introduction}

An important challenge in the implementation of quantum information protocols and in quantum simulation is to create strong, long-range, tunable and switchable interactions. Rydberg atoms have exaggerated properties, such as very large electrical polarizabilities and (induced) dipole moments \cite{Gallagher}. These characteristics make them very good candidates as mediators of the needed interactions, and are the reason why they are now being widely pursued as systems for quantum information science using various approaches, including cavity quantum electrodynamics \cite{cavityqed}, trapped ions \cite{ions1} and neutral atoms \cite{Saffman1,transistor1,transistor2}.  Among these approaches the combination of atom chips with neutral atoms offers unique opportunities to study quantum-degenerate gases \cite{atomchipbook} with the advantage of having a compact system that allows the efficient manipulation of quantum gases. Due to their sensitive response to electric fields, Rydberg atoms are also used as a tool for electrometry \cite{abel2011electrometry}. One of the disadvantages of using Rydberg atoms in an atom chip experiment is the presence of spatially inhomogeneous electric fields \cite{hermann1,surface1,surface2}. These fields are produced e.g. by adsorbates (deposited) on the surface of the chip \cite{surface1,surface2,rbsurface} or by a voltage drop across current-carrying wires on the chip. Due to the large polarizability of Rydberg atoms and their proximity to a surface, the coherence of the excitation will be limited by these stray electric fields. This makes the observation of Rydberg-mediated interactions in such systems a challenge. A detailed characterization of the mentioned stray electric fields is therefore crucial. 

Here, we employ two-photon Rydberg Stark-Zeeman spectroscopy of a cloud of magnetically trapped ultracold $^{87}$Rb atoms ($T\sim2.5$ $\mu$K), to characterize the local electric fields from an atom chip surface ($\sim100$ $\mu$m distance). An additional auxiliary electric field is generated at the location of the atoms by applying a voltage to an on-chip wire adjacent to the magnetic trapping wire. We show that it is possible to characterize the two uncompensated vector components of the local electric field (the third component is compensated by the auxiliary electric field) by a careful analysis of the measured Stark-Zeeman spectra of $S$ and $D$ states. The results are consistent with the calculated structure of Rydberg states in combined magnetic and electric fields. We also characterize the electric field gradient along the long axis of the cloud. 

This paper is structured as follows. In section \ref{SECTIONII} we summarize our calculations of $D$ state Stark-Zeeman maps for different field configurations and show how these are affected by residual fields in different directions. The experimental setup and data acquisition procedure are explained in section \ref{SECTIONIII}, followed by the main experimental results and a comparison with the calculations described in the previous section.  A summary of the main results is given in section \ref{SECTIONIV}.

\section{Stark map simulations}
\label{SECTIONII}

To compare our experimental results to theory, we need to calculate the Rydberg energies and eigenstates in combined electric and magnetic fields, with an \textit{a priori} unknown angle between the two. To this end, we first calculate field-free radial Rydberg wave functions, then include the electric field by calculating the matrix elements of the electric field operator and diagonalizing the result to obtain Stark eigenstates and eigenenergies, and finally take into account the magnetic field as a small perturbation in the Hamiltonian, to obtain combined Stark-Zeeman maps. We have verified that the results of our calculations are consistent with results from open-source Rydberg calculator packages \cite{rycalculator,rycalculator2} that have recently become available. 

In more detail, the calculation of the desired Stark-Zeeman maps starts from the binding energies $E_{b}=-R/\left[2(n-\delta)^{2}\right]$ of the field-free Rydberg states, given by the experimentally determined quantum defects $\delta_{n,l,j}$ of $^{87}$Rb for $S$ and $D$ states \cite{defects}, for $P$ states \cite{defects1}, $F$ states \cite{defects2} and $G$ states \cite{defects3} and the reduced Rydberg constant $R$ for $^{87}$Rb. Here $n$ is the principal quantum number, $l$ the orbital angular momentum, and $j$ the total angular momentum of the valence electron. For $l>4$ the quantum defect is negligibly small and we set it to zero. Under field-free conditions, the Rydberg wave functions separate into a product of a radial wave function and a remaining function describing the electron spin and angular part of the wave function. 
The radial wave functions $\psi_{n,l,j}(r)$ are obtained numerically using the Numerov method \cite{numerov1,numerov2}, integrating the radial Schr\"odinger equation inward starting from the classically forbidden outer region of the Coulomb potential with an energy given by the above quantum defects, and with a variable step size adapted to the changing spatial oscillation frequency of the wave function \cite{kleppner}. The spin and angular functions are obtained from the standard angular momentum algebra using the Wigner-Eckart theorem \cite{WignerE}.
In an applied electric field, $m_{j}$ (corresponding to the projection of the angular momentum onto the electric field direction) remains a good quantum number. The energies and eigenstates depend on $m_{j}$ and can be obtained for each $m_{j}$ separately. To this end, a set of all states with energies around the energy of interest is selected (typically $\sim1000$ states) for each $m_{j}$, and a finite-size matrix is set up with the diagonal elements given by the (field-free) energies and the off-diagonal elements given by the matrix elements of the electric field operator. The latter are obtained using the above radial wave functions for the radial matrix elements and the Clebsch-Gordan coefficients for the angular part. The resulting matrix is diagonalized yielding a Rydberg Stark map of energies and eigenvectors for each $ m_{j}$ as a function of electric field strength $E$.

\begin{figure}
\centering
\includegraphics[width=0.48\textwidth]{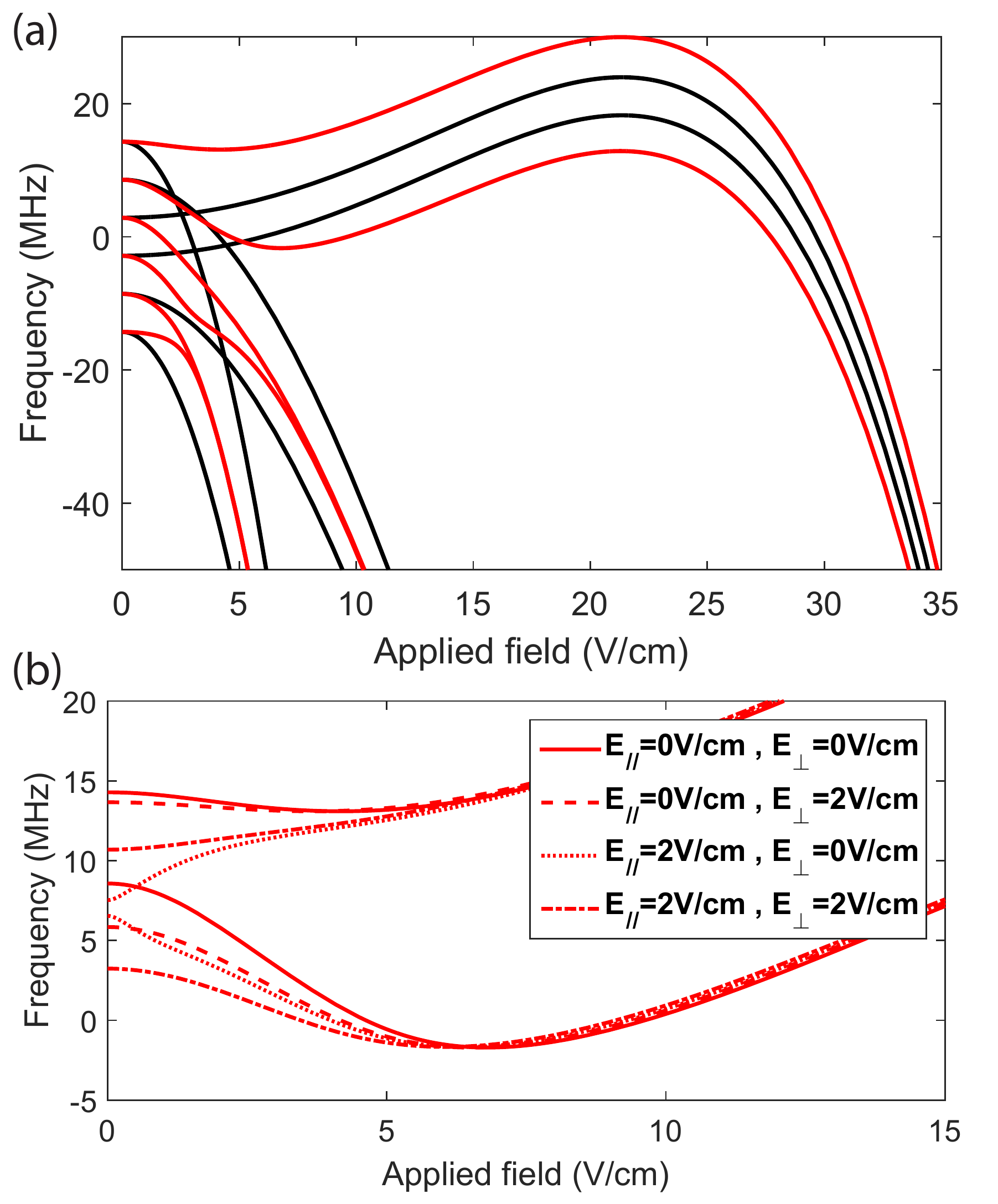}
\caption{(a) Calculated Stark map of the $28D_{5/2}$ state of $^{87}$Rb in a magnetic field of $B=3.4$ G. The red and black lines correspond to the case when the magnetic field is perpendicular and parallel to the electric field, respectively. In the $\vec{E}\perp\vec{B}$ configuration, for the two uppermost states the initial downshift is followed by an increase in energy, thus creating an initial dip in the Stark map. This is an additional feature compared to the $\vec{E}\parallel\vec{B}$ case and it can be used to extract information on different electric field components. (b) Detail of the shift of the two uppermost sublevels for various stray field configurations in the case where the applied electric field $\vec{E}_{ap}\perp\vec{B}$. This illustrates the sensitivity of the Stark-Zeeman map to different field configurations, see text for details.} 
\label{Starkmapallsublevels}
\end{figure}

To account for magnetic fields of a few Gauss, we limit ourselves to a single value for $n$, $l$ and $j$ (consistent with our magnetic trap) and the corresponding set of $(2j+1)$ basis states distinguished by their value of $m_{j}$ (with the quantization axis along the electric field direction). The corresponding eigenenergies $E(n,l,j,m_{j})$ yield a (diagonal) Stark Hamiltonian matrix within this manifold. Within this basis set the Zeeman Hamiltonian (with the magnetic field at an angle with respect to the electric field) is added as a perturbation, and the resulting matrix is again diagonalized to yield a Stark-Zeeman map of eigenvalues and eigenstates. The results of such a calculation are shown in Figure \ref{Starkmapallsublevels}.

From these calculations it is possible to analyze the character of each sublevel after a projection of the eigenbasis on the $B$ field direction. At zero electric field each sublevel is defined by the Zeeman shift. In Figure \ref{Starkmapallsublevels}(a) this means that each curve represents $m_{j}=5/2,3/2,1/2,-1/2,-3/2,-5/2$ from top to bottom, respectively. In the case in which $\vec{E}\parallel\vec{B}$ (black curves in Figure \ref{Starkmapallsublevels}(a)) the analysis is trivial because the quantization axis is unambiguous hence each sublevel maintains its own original character at any electric field value. As a result the Stark shift (which depends only on the absolute value $\lvert m_{j}\rvert$) and Zeeman shift are simply additive. The analysis is more complicated for the case when $\vec{E}\perp\vec{B}$. As the electric field is increased the character of each state changes, for instance the highest energy state $m_{j}=5/2$ (along $\vec{B}$) changes its character to a superposition of $m_{j}=-1/2$ and $m_{j}=+1/2$ (along $\vec{E}$) at higher fields. The latter is true already before the first maximum in energy (at an electric field of $\sim20$ V/cm). For intermediate values of the field there is also an admixture of the $m_{j}=3/2$ state. This analysis is important when setting the polarization of our lasers for the Rydberg excitation in order to select which state we excite in view of the selection rules. For the sake of simplicity we will use the zero-electric-field labeling (ZEL) to identify the various  sublevels throughout this paper, even when the electric field has non-zero value. 

From Figure \ref{Starkmapallsublevels}(a), we see that in the case of perpendicular fields (red lines), $\vec{E}\perp\vec{B}$, the energy of the two uppermost magnetic sublevels first decreases, followed by an increase of energy at higher fields, thus creating a dip in the Stark map. The size of this initial dip is very sensitive to additional (stray) electric fields. In Figure \ref{Starkmapallsublevels}(b) we illustrate this by showing four different configurations in which we varied the contributions of residual fields between $0$ V/cm and $2$ V/cm. These contributions were taken parallel to the magnetic field ($E_{\parallel}$) and perpendicular to both the applied electric field and magnetic field ($E_{\perp}$), a sketch of the field geometry is shown in Figure \ref{setupCHIP}(a)). A stray field in the parallel direction of $2$ V/cm already shows a total disappearance of the local minimum for the top state. More generally, the precise shape of the Stark-Zeeman map of the top two states is very sensitive to the strength of the electric field components $E_{\parallel}$ and $E_{\perp}$, enabling the characterization of these fields. The role of the magnetic field is relevant because its value defines the energy difference between each magnetic sublevel at zero electric field, which is the Zeeman shift. Moreover, Figure \ref{Starkmapallsublevels}(b) shows that this energy difference varies as the electric field is increased, revealing the importance of the magnetic field configuration in the analysis of the data. Consequently, the combination of electric and magnetic fields is crucial for the full characterization of residual unknown stray electric fields.

\section{Experimental results}
\label{SECTIONIII}
\begin{figure*}
\centering
\includegraphics[width=0.75\textwidth]{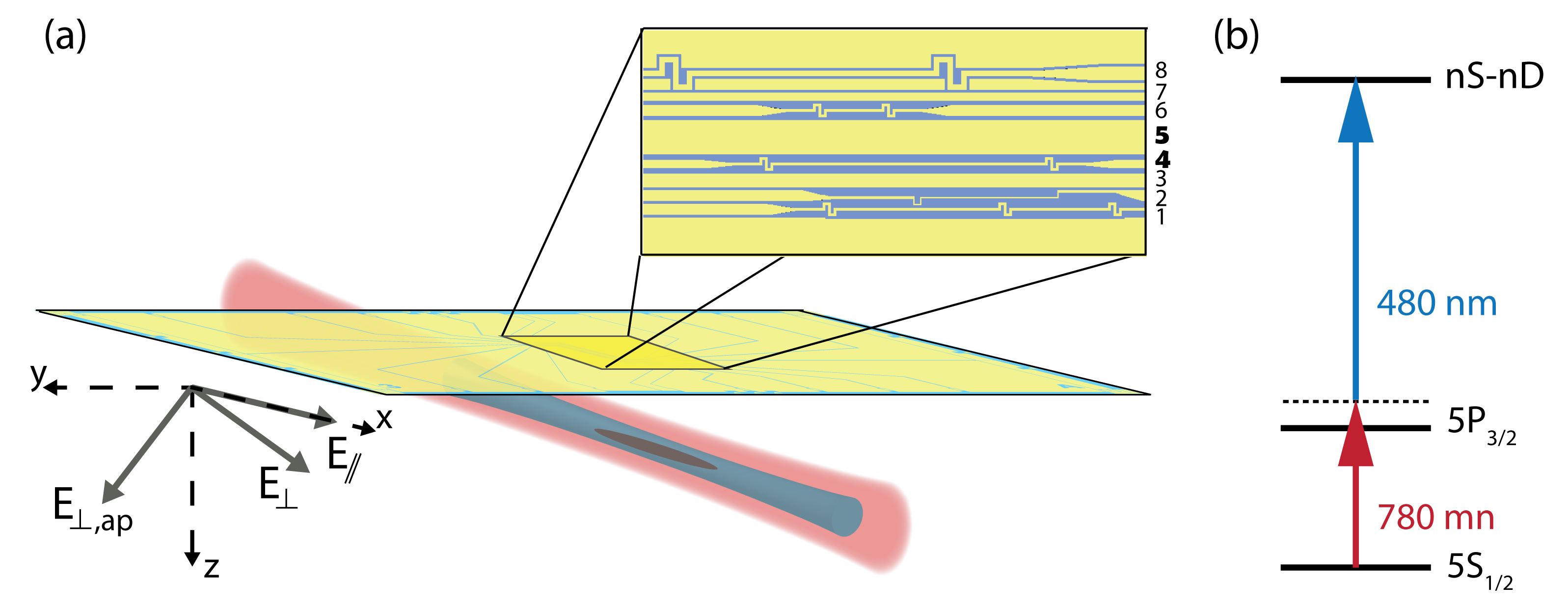}
\caption{(a) Sketch of the experimental geometry: We excite $^{87}$Rb atoms to a Rydberg level using a two-photon excitation scheme with two counter-propagating $780$ nm and $480$ nm lasers. The atoms are magnetically trapped in vacuum $\sim100$ $\mu$m below the surface of an atom chip. The numbering of the on-chip wires is indicated in the inset. The excitation is done along the axial direction of the cloud during magnetic trapping. We call the stray electric field along this direction the parallel component, $E_{\parallel}$. We apply a voltage to wire 4 that changes the field ($E_{\perp,ap}$) at the position of the atoms so we can generate Stark maps. The produced field is perpendicular to the $B$ field and lies in the $yz$ plane. The stray field in the third direction, labeled $E_{\perp}$, is in the $xz$ plane and orthogonal to $E_{\perp,ap}$. (b) Laser scheme for the Rydberg excitation. The intermediate state detuning from the $5P_{3/2}$ state is $100$ MHz towards the blue. }
\label{setupCHIP}
\end{figure*}
The experimental setup employs an atom chip \cite{atomchipbook} to trap and cool a cloud of Rb atoms in vacuum at a distance of $\sim100$ $\mu$m from a set of microfabricated gold wires (see Figure \ref{setupCHIP}(a)). Our setup and methods have been described in detail elsewhere \cite{setupthesis1,vanEs,setupthesis2}. In brief, we start from a cloud of $^{87}$Rb atoms loaded in a mirror-magneto optical trap (mMOT) with the mirror formed by the atom chip, a patterned, $2$ $\mu$m thick, gold layer on a $16\times 25$ mm$^{2}$ silicon substrate. These atoms are then optically pumped to the $\ket{F=2,m_{F}=2}$ state and transferred to a magnetic trap where they are cooled down to $\sim2.5$ $\mu$K using RF-induced evaporative cooling. The magnetic trapping is done using a $125$ $\mu$m-wide on-chip Z-shaped wire (wire labeled 5 in Figure \ref{setupCHIP}(a)) carrying a current of $1$ A. This final trap is elongated (cigar-shaped), with trap frequencies  $\omega_{x}/2\pi=46$ Hz and $\omega_{y,z}/2\pi=860$ Hz. The bottom of the trap is at $B=3.41$ G (corresponding to an RF frequency of $2.39$ MHz). After evaporation there are approximately $10^{4}$ atoms in the trap. At $T=2.5$ $\mu$K the calculated cloud size (FWHM) is $6.8$ $\mu$m in the radial directions and $127$ $\mu$m in the longitudinal direction. The temperature of the cloud is extracted from a time of flight (TOF) measurement in which the expansion of the cloud is measured as a function of the time after it was released. 

This setup has been extended with a two-photon Rydberg excitation scheme. We excite atoms from the $5S_{1/2}$ ground state to a Rydberg level via the intermediate state $5P_{3/2}$ using a $780$ nm infrared laser and a $480$ nm blue laser in a $7$ ms pulse during magnetic trapping. Both lasers are aligned along the long ($x$) axis of the magnetic trap and the lasers are frequency-narrowed and stabilized by locking them to a home-built reference cavity that is described in detail elsewhere \cite{homebuildcavity}. With this scheme we reach a linewidth $\lesssim10$ kHz for both lasers. For the Rydberg excitation we use a blue and infrared power of $90$ mW and $\sim0.5$ $\mu$W respectively. In order to increase the two-photon Rabi frequency the coupling blue beam is focused down to a waist of  $90$ $\mu$m. This leads to a blue Rabi frequency at full power of $\Omega_{c}\sim10$ MHz for the $5P_{3/2}-30S_{1/2}$ transition. The infrared beam waist at the position of the atoms is $520$ $\mu$m. The detuning to the intermediate $5P_{3/2}$ state is $100$ MHz towards the blue for the $780$ nm laser to reduce losses due to intermediate state scattering. Finally, after the Rydberg excitation pulse, the remaining ground state atoms are released and detected after a time of flight using absorption imaging.

By scanning the blue frequency across resonance and measuring the number of remaining atoms as a function of this frequency we obtain Rydberg loss spectra. In order to change the electric field at the position of the atoms we pulse a voltage during the Rydberg excitation. This voltage is applied to one of the on-chip wires (wire 4 in Figure \ref{setupCHIP}(a)) during the magnetic trapping/Rydberg excitation phase. 
\begin{figure}[ht]
\includegraphics[width=0.45\textwidth]{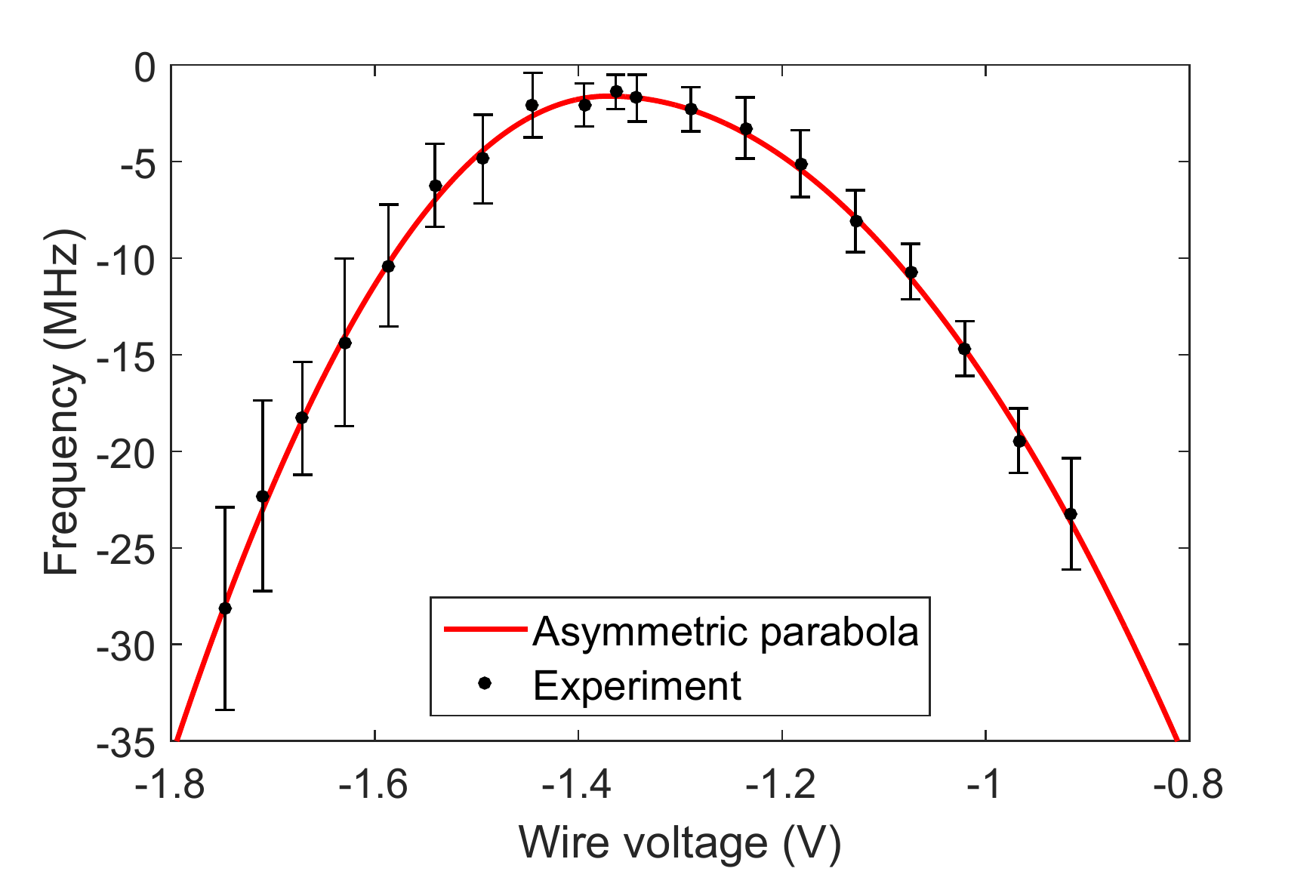}
\caption{Measured Stark shift and broadening for the $30S$ state. We use these measurements to calibrate the actual applied electric field we are generating at the position of the atoms. The plot shows the resonance position of the Rydberg signal at different applied voltages. The vertical bars are the measured FWHM of each feature obtained from a Gaussian fit. We fit two quadratic functions to the data in order to obtain the parameters in Eq.(\ref{voltagefield}), see the main text for details.}
\label{30SStarkmap}
\end{figure}
\subsection{Stark map of the $S$ state}
 In order to calibrate the relation between the applied voltage and the locally applied electric field at the position of the atoms we first measured a Stark  map of the $30S_{1/2}$ state by taking Rydberg loss spectra for various applied voltages. We set the polarization of the blue and infrared lasers to be $\sigma_{-}$ and $\sigma_{+}$, respectively. We fitted a Gaussian to each Rydberg spectrum in order to obtain the amplitude, position, and full width at half maximum (FWHM) of the loss feature. The resonance position and width of the Rydberg feature are plotted as a function of the applied voltage in Figure \ref{30SStarkmap}. The vertical bars represent the FWHM of each feature in the spectrum. As expected, in the presence of electric field gradients (see below, in Sec.\ref{subsectiongradients}), the feature gets broader as we increase the applied field. We found that the voltages on the on-chip wires are slightly asymmetric around the applied voltage on wire 4, $V_{ap}$. We attribute this behavior to a leakage current between the chip wires that is slightly asymmetric in the voltage difference between the wires. This produces an asymmetry of the shift in frequency around $\sim1.4$ V. To account for this we fit separate quadratic functions with a shared apex to the data in order to produce a voltage to field relation. They smoothly connect at the top, but have different curvatures. The resulting fits are shown in Figure \ref{30SStarkmap}. Using the known polarizability of the $30S$ state \cite{polarizabilities} we can extract a relation between the applied voltage and the field generated at the position of the atoms using
\begin{equation}
\Delta=-\frac{1}{2}\alpha_{30S}E^{2},
\end{equation}
where $\alpha_{30S}=1.39$ MHz/(V/cm)$^{2}$ is the polarizability of the $30S$ state and $\Delta$ is the energy shift of the Rydberg feature produced by the presence of an electric field $E$. Using these two different fits we obtained a field-voltage relation of the form
\begin{equation}
E_{ap}=c_{i}\left(V_{ap}-V_{0}\right).
\label{voltagefield}
\end{equation}
Here $E_{ap}$ is the applied field at the position of the atoms, and $c_{i}$ are two different coefficients depending on which side of the parabola the applied field is: $c_{1}=12.4$ cm$^{-1}$ and $c_{2}=16.4$ cm$^{-1}$ for $V_{ap}>-1.37$ V and $V_{ap}<-1.37$ V respectively. $V_{0}=-1.37$ V is the offset voltage we get for which the Stark shift is minimal. This voltage is consistent with the inferred voltage at the center of the trapping wire (wire 5 in Figure \ref{setupCHIP}(a)) produced by the current we send through it to generate the magnetic trap. Note that a small stray field along the direction of $\vec{E}_{ap}$ would lead to a small shift of $V_{0}$ in the Stark map. In contrast, stray fields in the other two directions (orthogonal to $\vec{E}_{ap}$) lead to a vertical energy offset $-\frac{1}{2}\alpha_{30S}(E_{\parallel}^{2}+E_{\perp}^{2})$ in the Stark map of the $S$ state that does not discriminate between $E_{\parallel}$ and $E_{\perp}$, and that would require an absolute frequency reference to calibrate.

\subsection{Stark map of the $D$ state}
\label{dstate}
\begin{figure}[ht]
\includegraphics[width=0.45\textwidth]{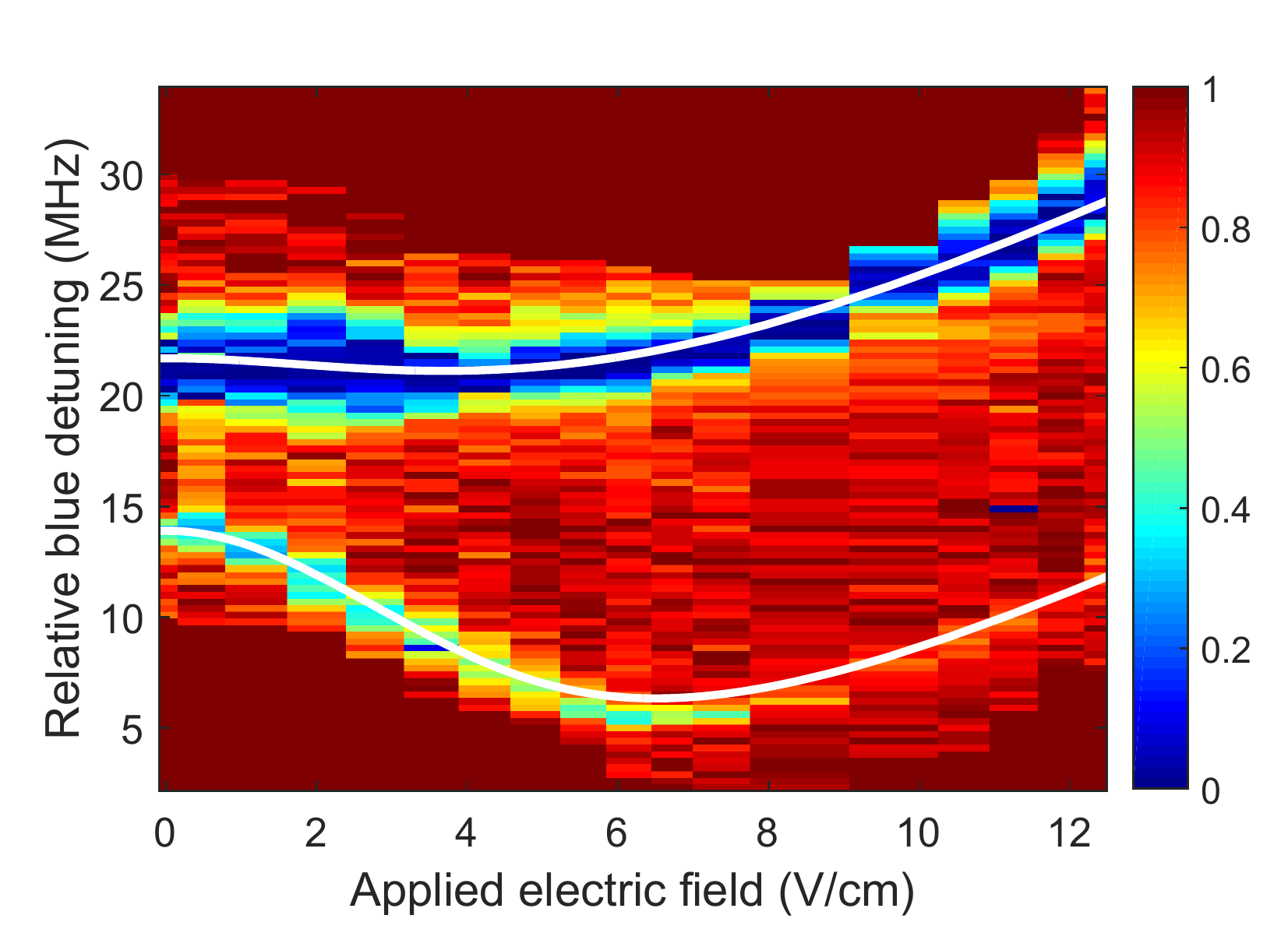}
\caption{$28D_{5/2}$ state Stark map of two different sublevels: The top and bottom curves are the $m_{j}=5/2$ and $m_{j}=3/2$ sublevels (ZEL) respectively. The color scale indicates the normalized number of atoms, the measurements were taken outside the dark red colored area. We deliberately saturated the upper sublevel in order to be able to see the other one, which has a much smaller coupling strength. The white lines are the resulting fitted curves for the resonance position of each sublevel where the parallel and perpendicular electric field are used as fitting parameters (from the fit we get $E_{\parallel}=0$ V/cm and $E_{\perp}=1.98$ V/cm, respectively. See main text for details).}
\label{28Dtwinpeaksfit}
\end{figure}

The knowledge of the applied electric field at the position of the atoms allows us to measure a calibrated Stark map of the $28D_{5/2}$ state. To do so we set the polarization of the blue and infrared beams to be linearly polarized, in such a way that we are able to see both the $m_{j}=5/2$ and $m_{j}=3/2$ states (ZEL). The resulting Stark-Zeeman map is shown in Figure \ref{28Dtwinpeaksfit}. We used Eq. (\ref{voltagefield}) to transform the $x$-axis from an applied voltage to an applied electric field at the position of the atoms. We fitted two Gaussians to each spectrum for a given applied field and extracted the main properties of the two features: amplitude, FWHM and position. The comparison of the resonance position obtained from the data with the calculated ones gives us values for the components of the stray field in both parallel, $E_{\parallel}$, and perpendicular, $E_{\perp}$,  direction. The analysis is based on the shape and the size of the initial dips in the Stark map (for applied fields $\leq 10$ V/cm) which strongly depends on the stray field configuration (see Figure \ref{Starkmapallsublevels}(b)). 
Considering the values obtained for the electric field components we can estimate to which extent the FWHM of each feature is limited by field gradients. 

From the Stark-Zeeman map in Figure \ref{28Dtwinpeaksfit} we already see the initial dip in the energy shift of the resonance of both states so we can immediately set an upper limit to the field in the parallel direction of $E_{\parallel}\lesssim 2$ V/cm. 
The coupling strength of the $m_{j}=3/2$ state is much weaker than that of the $m_{j}=5/2$ (ZEL) so we have to strongly saturate the latter in order to see both at the same time. The weak feature has a much more prominent dip as a function of electric field than the strong one. Therefore, the simultaneous use of the resonance positions of both states results in a more accurate analysis of the data. However, since the $m_{J}=5/2$ state is strongly saturated, it is not possible to extract an accurate value for the resonance position of the Rydberg feature. In order to get a better data set for this state we set the blue and infrared beam to be $\sigma_{+}$ polarized so only the transition to the $m_{j}=5/2$ state (ZEL) is allowed by selection rules. Because this state has a strong coupling it is possible to investigate its behavior also at higher electric fields without losing the signal. The resulting Stark-Zeeman map is shown in Figure \ref{28DStarkmap}(a). These data are combined with the previously measured $m_{J}=3/2$ state (ZEL) data, and fitted to the calculated Stark-Zeeman map (Figure \ref{Starkmapallsublevels}(b)) with the residual electric field in the parallel and perpendicular directions as fitting parameters. For the fit we use data for fields $E_{ap}<15$ V/cm only, from which we obtain $E_{\parallel}=0.0\pm0.2$ V/cm and $E_{\perp}=1.98\pm0.09$ V/cm\footnote{The stated uncertainties in the fields, $\Delta E_{\perp}$ and $\Delta E_{\parallel}$, are calculated assuming that the reduced chi-square value of the model is one, $\chi_{r}^{2}=1$. Thus, $\Delta E_{\perp}$ and $\Delta E_{\parallel}$ are obtained by allowing the chi-square value, $\chi^{2}$, to increase by one from its minimum value. The more usual approach of extracting the uncertainty from the inverse covariance matrix on basis of the curvatures at the minimum of $\chi^{2}$ is not applicable because the dependence of $\chi^{2}$ on $E_{\parallel}$ was found to be nonquadratic \cite{press1987numerical}.}. The resulting fitted curves are shown on top of the measured Stark-Zeeman map as white lines in Figures \ref{28Dtwinpeaksfit} and \ref{28DStarkmap}(a). 

Although the resulting fitted curves show excellent agreement with both sublevels at low field values, $E_{ap}<10$ V/cm, there is a discrepancy at high field values, $E_{ap}\approx20$ V/cm (see Figure \ref{28DStarkmap}(a)). We have verified that the frequency difference between the observed bump at $E_{ap}=20$ V/cm and the observed dip at $E_{ap}=5$ V/cm is reproducible and not affected by slow drifts in the experimental frequency calibration. This was done by measuring the frequency of the observed bump at $E_{ap}=20$ V/cm and the observed dip at $E_{ap}=5$ V/cm directly after one another and confirming their vertical spacing in Figure \ref{28DStarkmap}(a). In this regard it is noteworthy that one of the open-source packages \cite{rycalculator} actually yields a lower value of the peak at $20$ V/cm (about $1$ MHz) than our own calculations and those of \cite{rycalculator2} (albeit in the absence of magnetic fields, so a direct comparison with our experimental data is not possible). Another source for a reduction in the peak frequency shift at $20$ V/cm is the differential AC-Stark shift of the Rydberg level as the character of the state changes with increasing applied electric field. We calculated this to give a modest contribution, a downshift in frequency  of $\approx 0.5$ MHz at $E_{ap}=20$ V/cm. In short, the discrepancy between our calculation and the experimental results around $E_{ap}=20$ V/cm may be due to a combination of factors, and is still under investigation. This discrepancy does not affect the determination of $E_{\parallel}$ and $E_{\perp}$ because it is only visible at high field values ($E_{ap}>15$ V/cm), where small residual fields do not play a role.  

\subsection{Electric field gradients}
\label{subsectiongradients}
\begin{figure}[ht]
\includegraphics[width=0.5\textwidth]{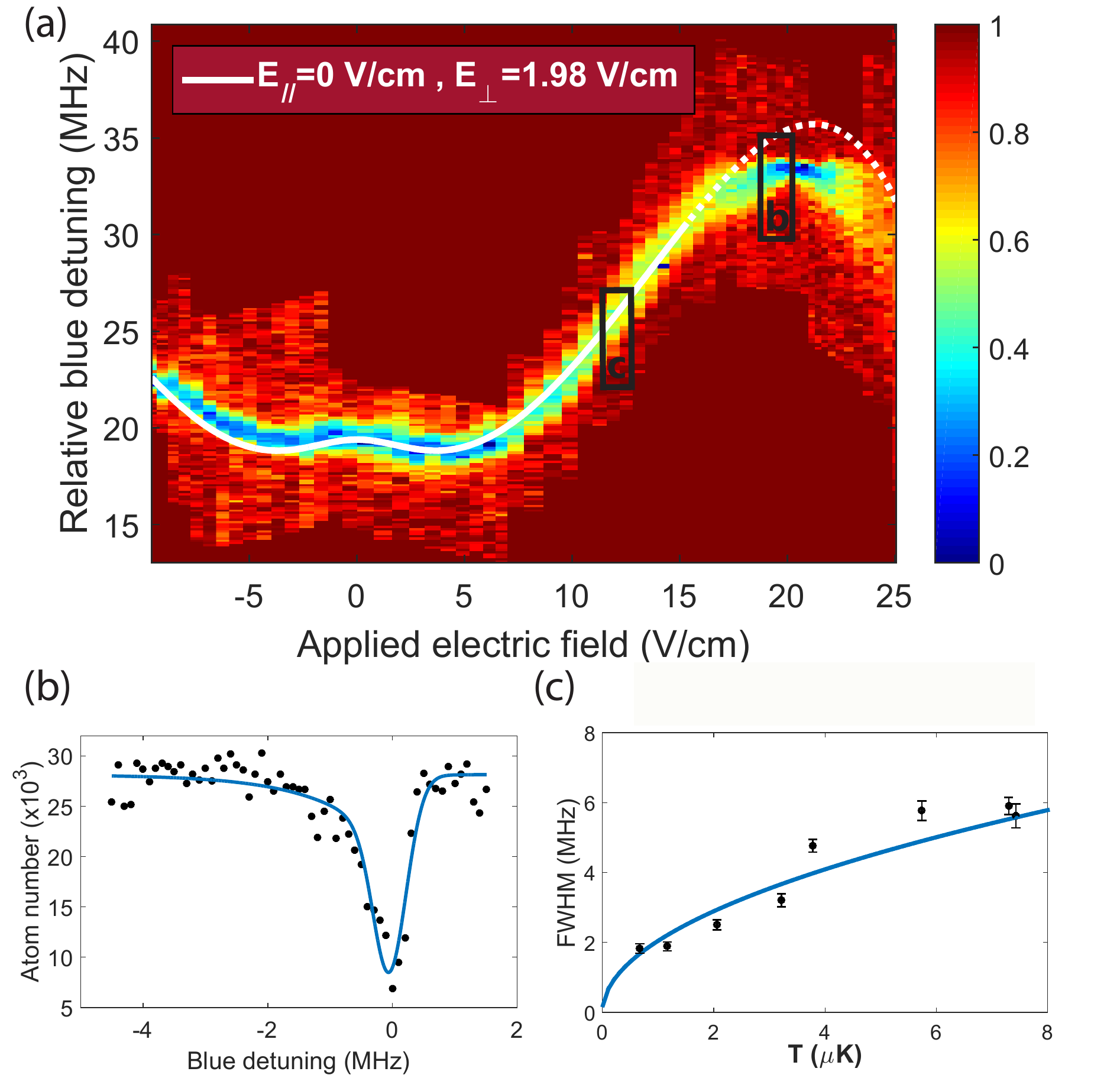}
\caption{Rydberg loss spectroscopy of the $28D_{5/2}$ state for different applied fields. (a) Shows the top state in Figure \ref{Starkmapallsublevels}(a) considering the voltage to field conversion. The color scale indicates the normalized atom number in each spectrum, measurements are performed outside the uniform dark red colored area. The dashed white line shows the resulting fitted curve, obtained using the data within the area of the solid white line only (below $15$ V/cm). (b) At $E_{ap}\sim19$ V/cm the Rydberg feature is asymmetric due to the presence of field gradients. Therefore a skewed Gaussian (solid curve) is fit to the data. (c) Shows the measurement of the FWHM of the loss feature at different temperatures at an applied electric field of $E_{ap}\sim12$ V/cm where the Rydberg feature is symmetric. }
\label{28DStarkmap}
\end{figure}

We now focus on possible gradients in the electric field as a source of broadening in the experimental spectra. Already from the calculations we can conclude that in the areas where the Stark shift has an extremum as a function of field the Rydberg loss feature will become asymmetric. This is due to the quadratic character of the energy shift around the extremum and the finite size of the cloud, which means that different parts of the cloud sample different values of the electric field. Figure \ref{28DStarkmap}(b) shows an example of a spectrum taken at $E_{ap}=19$ V/cm, where the effect of gradients in the Rydberg spectrum is enhanced due to the high curvature of the frequency-field relation. From this spectrum it is possible to see the asymmetry of the signal. This observed asymmetry supports the assumption that the broadening of the Rydberg feature is caused by the presence of field gradients at the position of the atoms. 
 
The above argument implies that the Rydberg feature is symmetric in the field regime where the energy shift is linear. This linear regime is a characteristic that is unique for $D$ states and that can actually be used to extract an approximate value of the electric field gradient in the volume of the atoms. 
To extract this value from the data we assume a Gaussian density distribution along the long direction of the cloud, $n(x,T)\propto\exp\left[-x^{2}/2 \sigma_{x}(T)^{2}\right]$, with a size set by the temperature: $\sigma_{x}(T)=\sqrt{k_{B}T/m\omega_{x}^{2}}$, where $k_{B}$ is the Boltzmann constant, $m$ the atomic mass and $\omega_{x}$ the trap frequency in the long direction. 

The cloud samples different values of the field, which are set by the gradient, $g$, and the size of the cloud via the expression $\delta E(x,g)=g x$. At the same time these fields are related to the linear shift in energy via $\Delta(x,g)=a \delta E(x,g)$ where $a=1.44$ MHz/(V/cm) is the slope we get when we fit a line to the part of the Stark shift of the upper state of the $28D_{5/2}$ manifold that has a linear behavior. The FWHM of the Rydberg spectrum depends on the field gradient, temperature (because it sets the size of the cloud) and the zero-field linewidth. The latter corresponds to the linewidth of the spectrum set by any other mechanism of broadening besides electric field effects. 
In order to extract the gradient, we set the applied electric field to $E_{ap}\sim12$ V/cm (where the shift is linear with the field) and vary the temperature by changing the final RF frequency in the evaporative cooling process. Finally the FWHM of the Rydberg loss spectrum is plotted as a function of temperature, the results are shown in Figure \ref{28DStarkmap}(c). We refrained from using a fitting function that includes a finite value for the FWHM at $T=0$ K, corresponding to the previously mentioned zero-field linewidth. This is not necessary in our fit due to the lack of data points for low temperatures (limited by signal to noise ratio because of the low atom number), further it does not affect the estimate of the gradient because this is mainly set by the slope of the fitted curve. From our data we obtained a field gradient of $g=179\pm9$ V/cm$^{2}$ along the long direction of the cloud. This is the value for a field gradient that explains the broadening of the Rydberg feature. In particular, it explains the increase of the FWHM for higher temperatures shown in Figure \ref{28DStarkmap}(c). Interestingly, it also allows a comparison with the maximum widths observed in the $S$ state spectrum of Figure \ref{30SStarkmap}. When the applied field dominates over the residual fields, the width of the spectrum is sensitive to the same electric-field gradient we determined above. The expected FWHM width of the spectrum when the applied field dominates is then $\alpha_{30S}E_{ap}g\delta x$, with $\delta x$ the FWHM length of the cloud. This yields a FWHM of $9$ MHz for a wire voltage of $-1.75$ V in Figure \ref{30SStarkmap}, consistent with the measured width.

Another reason that might explain the difference between the model and the experimental data at low temperatures is the omission of density effects. At low temperatures there is an important increase in density which means Rydberg-mediated interactions might be playing a role in the observed width. Our model does not consider this. However, the value of the gradient is set mainly by the slope of the curve which is set by higher-temperature data ($T>2$ $\mu$K). In this area, the dominant mechanism of broadening is the effect of gradients, particularly along the long direction of the cloud. The investigation of Rydberg-mediated interactions and collective effects at lower temperatures and higher densities is the subject of further work, and is beyond the scope of the present paper.\\

\section{Conclusions and outlook}
\label{SECTIONIV}
We characterized the vector components of the stray electric field close to the chip surface in our atom chip experiment by comparing calculated $S$ and $D$ Stark-Zeeman maps with experimental data.
We applied a voltage to one of the on-chip wires in order to generate a field at the position of the atoms. The field is characterized using a measured $S$ state Stark-Zeeman map, which is then used as a tool to calibrate the field axis of the $D$ state Stark-Zeeman map. The energy shift of $D$ states when applying an electric field is non trivial and has an initial dip in energy. The minimum of this dip, and therefore the change in curvature, strongly depends on the values of stray electric fields along different directions. The values of these field components are obtained by fitting the data to the simulated Stark-Zeeman map. We obtained values for the residual fields that are relatively small, namely $E_{\parallel}=0.0\pm0.2$ V/cm and $E_{\perp}=1.98\pm0.09$ V/cm. The total stray field is about a factor of two lower than the observed at the same distance ($100$ $\mu$m) under similar conditions using electromagnetically induced transparency \cite{surface1}. This reduction may be due to the somewhat elevated temperature of the current carrying gold wire in our system, which should lead to reduced Rb coverage. We also made use of the linear response of $D$ states to applied electric fields. In this region the Rydberg signal is symmetric and the increase in linewidth as we increase the size of the cloud is related to the field gradient at the position of the atoms. The value of the gradient allows us to set a lower limit for the linewidth of the Rydberg signal setting the stage for future research of Rydberg-mediated interactions in our setup. 

On one hand the elongated character of our cloud sets a limit for the linewidth we can observe due to different parts of the cloud sampling different fields but on the other hand it has been demonstrated that the $1$D character has advantages when observing Rydberg-mediated interactions \cite{dressing2}. From this research we can conclude that atom chip experiments are a promising tool for the study of Rydberg systems. The level of on-chip control over the electric fields demonstrated here is promising for the observation of Rydberg blockade and Rydberg-mediated interactions in ultracold gases trapped on a chip, which is relevant for  quantum information science and technology in integrated and compact systems. A next generation of chips is being designed in which the stray electric fields along $E_{\parallel}$ and$E_{\perp}$ can be compensated using electrodes placed as an extra layer on top of the chip. The characterization of the fields in the new setup will be done using the electrometry process described in this paper.
\\

\section{Acknowledgments}
This work was financially supported by the Foundation for Fundamental Research on Matter (FOM), which is part of the Netherlands Organisation for Scientific Research (NWO). We also acknowledge financial support by the EU H2020 FET Proactive project RySQ (640378).

 \bibliography{References}

\end{document}